\documentstyle[a4,12pt]{article}
\begin{document}
\rightline{DAMTP/97/57}
\rightline{SPhT/97/57}
\vskip .2 cm
\centerline{\bf EXPONENTIALLY SMALL COUPLINGS}
\vskip .5 cm
\centerline{\bf AND}
\vskip .5 cm
\centerline{\bf THE HIERARCHY PROBLEM}
\vskip .5 cm
\centerline{by}
\vskip .25 cm
\centerline{{Ph. Brax}\footnote{brax@spht.saclay.cea.fr}}
\vskip .25 cm
\centerline{Service de Physique Th\'eorique}
\vskip .25 cm
\centerline{CEA-Saclay 91191 Gif/Yvette cedex France}
\vskip .5 cm
\centerline{and}
\vskip .5 cm
\centerline{{Neil Turok}\footnote{n.g.turok@amtp.cam.ac.uk}}
\vskip .25 cm
\centerline{Department of Applied Mathematics and Theoretical Physics}
\vskip .25 cm
\centerline{Silver Street Cambridge CB39EW UK}
\vskip .5 cm
\leftline{\bf Abstract:}
\vskip .5 cm
We propose a stringy mechanism whereby a large hierarchy between symmetry breaking scales is generated. This mechanism is based upon the existence of a fifth dimension compactified on a segment. We focus on a simple supersymmetric model with one massless Higgs field in the $3$ of $SU(3)$ and another one in the $\bar 3$ on each four dimensional end point of the fifth dimension.  Along supersymmetric flat directions the gauge symmetry is broken down to $SU(2)$ due to the vacuum expectation value of the Higgs fields on one of the end points. The remaining massless mode on the other end point acquires a potential due to a massive five dimensional state propagating between the end points. This potential breaks the $SU(2)$ symmetry at an exponentially suppressed scale compared to the $SU(3)$ breaking scale. The suppression factor depends exponentially on the mass $M$ of the massive state and the length $\pi R$ of the fifth dimension . For reasonably large values of the length scale $R$ one can achieve a factor of order $M_{W}/M_{GUT}$.
 
\vfill\eject
The hierarchy problem, namely the large gap between the weak scale where the standard model is successful and the Planck scale where gravitational effects are  relevant, is a well-known puzzle.  Ordinary field theories do not explain why the weak scale is small and protected from large radiative corrections. On the other hand supersymmetric extensions of the standard model  provide a natural framework within which radiative corrections are tamed due to the cancellation of bosonic and fermionic loops$^{[1]}$. However supersymmetry has to be broken at a scale larger than the weak scale as superpartners have not been observed. The breaking of supersymmetry is generally supposed to occur at a scale close to the Planck scale where gravitational effects in a hidden sector give masses to the superpartners$^{[2,3]}$. The electroweak breaking is then induced by radiative corrections in the Higgs sector$^{[4]}$. Typically this mechanism requires that the gravitino mass $m_{3\over 2}$ has to be of the order of the weak scale in order to reproduce the observed spectrum of particles$^{[5,6,7,8,9]}$. This replaces the hierarchy problem by the problem of the smallness of the soft supersymmetry breaking terms. 
In order to avoid this loophole one can look for another mechanism for the electroweak symmetry breaking. We suggest that the origin of the electroweak symmetry breaking may be more geometrical and linked to the existence of extra dimensions$^{[10]}$.

 It has been recently shown by Horava and Witten$^{[11]}$ that the large coupling limit of the heterotic string is given by the compactification of the eleven dimensional M-theory. At the level of the low energy effective supergravities, this amounts to compactifying 11d supergravity on a segment. On each of the end-points of the segment, i. e. in ten dimensions, the gauge degrees of freedom reproduce one of the $E_8$ factors of the heterotic string. Upon compactification  to five dimensions on a Calabi-Yau threefold $^{[12,13,14,15,16]}$the theory possesses the unusual features of having an extra dimension with two end-points. Inspired by the example of M-theory we propose a simple stringy situation where the existence of a segment in the fifth dimension plays a crucial role in generating large scale hierarchies.
 We assume that the four dimensional theory   is supersymmetric after compactification. We suppose that the gauge degrees of freedom have been reduced  to a product $SU(3)\times SU(3) \times U(1)$. This can be envisaged by breaking of the $E_8$ gauge symmetry.  We choose to study the Higgs sector  
comprising two Higgs fields $\phi_i$ and $\bar\phi_i$ in the $3$ and $\bar 3$ of the second $SU(3)$ on each of the end-points $i=1,2$ of the segment. We also require the existence of  a massive scalar state of mass $M$  propagating in  five dimensions. This state belongs to the adjoint representation the gauge group and couples to the Higgs fields. 
We demonstrate that these minimal requirements are enough to induce large scale differences between the breaking of $SU(3)$ and the subsequent breaking of $SU(2)$.

Let us first define our model more precisely. We consider a higher dimensional string theory whose fifth dimension is compactified on the quotient $S^1/Z_2$ where the $Z_2$ symmetry acts as $x_5\to -x_5$. In particular the fifth dimension is isomorphic to a segment $I$ of length $\pi R$ where $R$ is the radius of the circle $S^1$. Each of the end points of the segment correspond to a copy of a four dimensional Minkowsky space. The compactification on such an orbifold$^{[17,18,19]}$ leads to different sectors of the theory. The untwisted sector corresponds to strings wrapping  around the circle $S^1$. This is for instance the case of the Kaluza-Klein winding modes satisfying $X\equiv X + 2\pi nR$ of masses $M_n={nR\over 2}$. Here and below we adopt units in which the string tension $T={1\over 4\pi}$. The twisted sectors are identified with strings closed up to the $Z_2$ symmetry $X\equiv -X+2\pi nR$. The corresponding states are localised at one of the end points of the fifth dimension. We assume that the Higgs fields belong to the twisted sector.
 The gauge fields are coupled to two pairs of Higgs fields living on each end point. The superfields $\phi_1$ and $\bar\phi_1$ are fixed at $x_5=0$ while  $\phi_2$ and $\bar\phi_2$ are fixed at $x_5=\pi R$. The Lagrangian contains  the four dimensional term
\begin{equation}
L=\int d^4 x (\int d^4\theta (\phi_i^*e^V\phi_i+\bar\phi_i^*e^{-V}\bar\phi_i)+{1\over 4\pi g^2}\int d^2 \theta W^2)
\end{equation}
where $V$ is an $SU(3)$ vector superfield and $W$ the chiral superfield containing the field strength.
The chiral superfields are evaluated at $x_5=0$ for $i=1$ and $x_5=\pi R$ for $i=2$.   
The  scalar potential reads
\begin{equation}
V ={1\over 2}g^2\vert D^a\vert ^2 
\end{equation}
where $g$ is the  value of the gauge coupling constant and the $D$ terms are
\begin{equation}
D^a=\phi^*_{iu}T^a_{uv}\phi_{iu}-\bar\phi^*_{iu}T^a_{uv}\bar\phi_{iv} 
\end{equation}
 where $T^a$ are the generators of $SU(3)$.
As supersymmetry is not broken the vacua are determined by $D^a=0$.  Using gauge and flavour  transformations one can always write
$$
\phi_{iu}=\bar\phi_{iu}=
\left ( \begin{array}{ccc}
a& 0 &0\\
0& b&0
\end{array}\right )
$$
where $a$ and $b$ are real parameters. Notice that if $b=0$ the gauge group is broken down to $SU(2)$ whereas it is completely broken when $a$ and $b$ are non-zero. We will have to assume that $\phi_1$ and $\bar \phi_1$ acquire a non-zero vev in order to break the gauge symmetry from $SU(3)$ to $SU(2)$. The gauge group is broken on one of the end points, i. e. the Higgs fields on the second end point are not affected by the first step of symmetry breaking.  
The potential in the $b$ direction is completely flat.
This flat direction only concerns the fields on the second end point and is lifted due to the presence of  a massive particle in the bulk. The massless mode  localised on $x_5=\pi R$ picks up a non-vanishing potential due to the presence of a massive state between the end points. 

Consider the interaction of two twisted string states. They interact and emit an untwisted winding mode. If the winding of the resulting mode vanishes it corresponds to the emission of a massless gauge boson in the adjoint representation$^{[20]}$. Similarly the massive winding modes belong to the adjoint representation. These massive modes couple twisted states fixed at either the same or  at different fixed points. Let us focus on the lightest of these massive Kaluza-Klein states $\vert \xi>$ in the adjoint representation of $SU(3)$. The effect of this state is two-fold. It can either propagate between two points on the same boundary or travel from one to the other end point. The two processes have different consequences on the structure of the potential. The former gives rise to contact terms while the latter gives rise to exponentially suppressed four-point couplings. In a first quantised formulation the couplings arise from the propagator of the state $\vert \xi>$. Introducing the Green function 
\begin{equation}
G(x,y)=\sum_{n\in Z}\int_0^{\infty} e^{-{{tM^2}\over 2}}\int_x^{y+2\pi nR}{\cal D}x(t)e^{-{1\over 2}\int_0^t\dot x^2d\tau}
\end{equation}
on the circle, the propagator on the segment reads 
\begin{equation}
\Delta (x,y)={1\over 2}(G(x,y)+G(-x,-y))
\end{equation}
due to the $Z_2$ symmetry. 
\begin{figure}
\vspace{7cm}
\special{hscale=60 vscale=60 voffset=0 hoffset=10 psfile=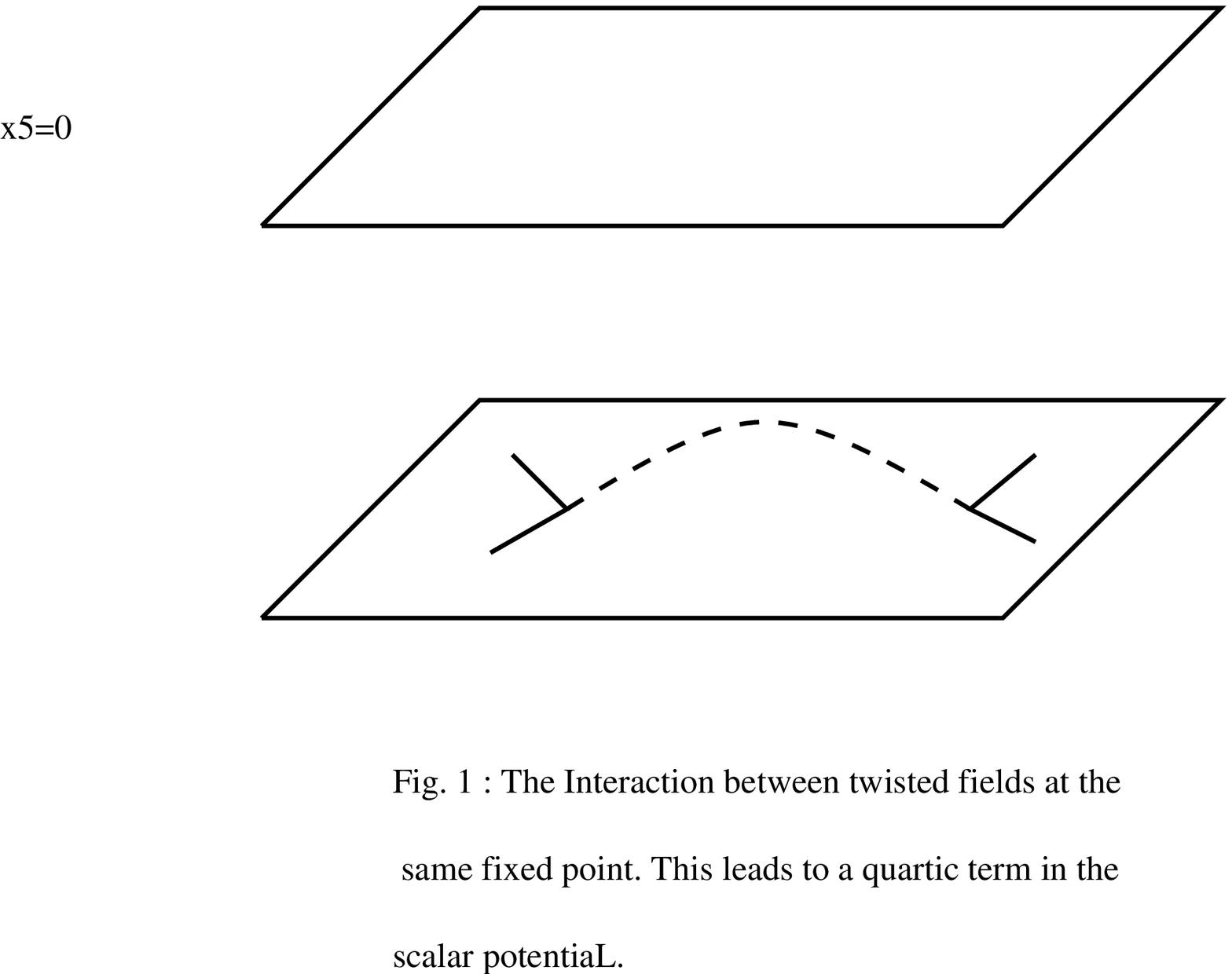}
\end{figure}
The contact  amplitude reads
\begin{equation}
{\lambda^2\over R}T^a_{ij}T^a_{kl}\Delta (0,0)
\end{equation}
leading to a contact term in the scalar potential
\begin{equation}
V_c={\lambda^2\over R}  \Delta(0,0)\vert (\phi_2^*+\bar\phi_2)T^a(\phi_2+\bar\phi_2^*)\vert^2
\end{equation}
where $\lambda$ is a dimensionless coupling constant and the factor of $R$ is introduced for dimensional reasons. 
The value of the contant $\Delta (0,0)$ can be deduced by a saddle point approximation leading to
\begin{equation}
\Delta (0,0)\sim {1\over M}{1\over 1-e^{-2\pi MR}}
\end{equation}
\begin{figure}
\vspace{7cm}
\special{hscale=60 vscale=60 voffset=0 hoffset=10 psfile=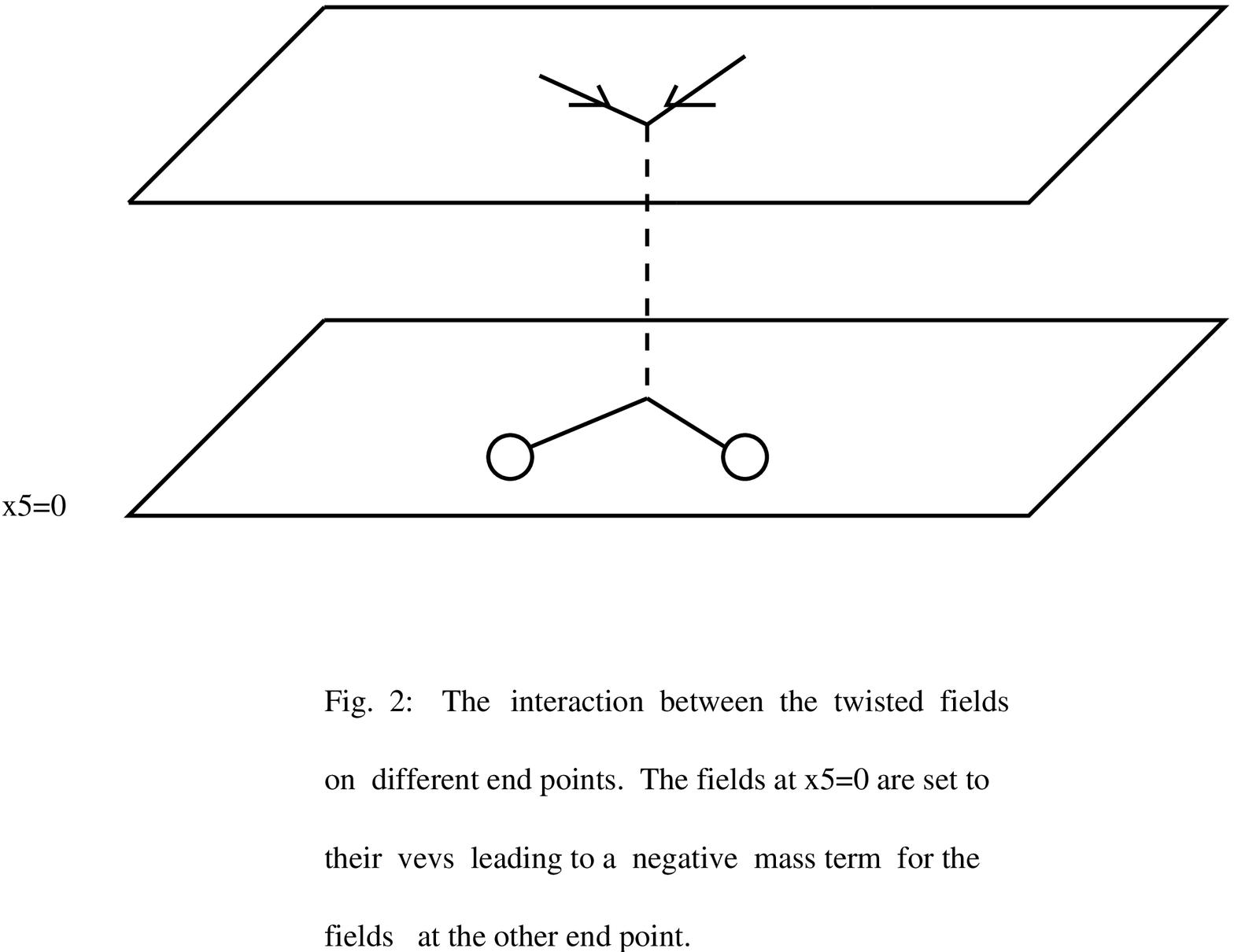}
\end{figure}
Similarly the massive state can couple the fields on the different end points with an amplitude
\begin{equation}
{\lambda^2\over R}T^a_{ij}T^a_{kl} \Delta (0,\pi R)
\end{equation}
This leads to an exponentially small coupling$^{[20]}$
\begin{equation}
V_e={\lambda^2\over R}\Delta (0, \pi R) ((\phi_2^*+\bar\phi_2)T^a(\phi_2+\bar\phi_2^*))
((\phi_1^*+\bar\phi_1)T^a(\phi_1+\bar\phi_1^*)) 
\end{equation}
where $\phi_1$ is equal to its vev.
The constant $\Delta (0,\pi R)$ can be evaluated by a saddle point approximation and leads to
\begin{equation}
\Delta (0,\pi R)=e^{-\pi MR}\Delta (0,0)
\end{equation}
The potential $V_e$ gives a quadratic term to the $b$ potential whereas the term $V_c$ gives a quartic term.  Using the Gell-mann generators of $SU(3)$ one can find the $b$ potential
\begin{equation}
V_b=\kappa b^2(2b^2- e^{-\pi MR}a^2)
\end{equation}
where 
$\kappa= {{16\lambda^2}\over 3R}\Delta (0,0)$
The induced potential possesses a negative mass term. This implies that the vev of $b$ is given by 
\begin{equation}
b={e^{-{{\pi RM}\over 2}}\over 2}a
\end{equation}
The $SU(2)$ gauge symmetry is broken at a scale exponentially small compared to the $SU(3)$ breaking scale. 

Let us now consider the fermion masses. We now  take into account the other $SU(3)\times U(1)$ factors of the gauge group. We suppose that the Higgs fields are not charged under $SU(3)\times U(1)$. This implies that the $D$ terms are not modified. Introduce the left quark fields $Q\in (3,\bar 3)$ of $SU(3)\times SU(3)$, the right conjugate quark fields $U\in (\bar 3,1)$ and $D\in (\bar 3,\bar 3)$. Similarly the left leptons are $L\in (1,\bar 3)$ and the right conjugate leptons $E\in (1,\bar 3)$. Notice that the $E$ and $D$ fields are triplets under the second $SU(3)$. We also need fields $D'\in (3,\bar 3)$, $E'\in (1,\bar 3)$ and $U'\in (\bar 3,1)$ in order to give a mass to the extra components of $E$, $D$ and $Q$. 
 This allows to write down a superpotential 
\begin{equation}
W=\lambda_U UQ_i\phi_{2i}+\lambda_{U'}U'Q_i\phi_{1i}+
\epsilon_{ijk}(\lambda_DD_i\bar\phi_{2j}Q_k+\lambda_E  E_i\bar\phi_{2j}L_k+ \lambda_{D'}\bar\phi_{1i}D_jD'_k +\lambda_{E'}\bar\phi_{1i}E_jE'_{k})
\end{equation}
where only the second $SU(3)$ indices are shown. Notice that after $SU(3)$ breaking along the flat direction the fields $U'$, $Q_1$, $D_{2,3}$, $D'_{2,3}$, $E_{2,3}$ and $E'_{2,3}$ become massive. They therefore decouple. Decomposing the $\bar 3=2+1$ under $SU(2)$ one gets the following superpotential 
\begin{equation}
W=\lambda_U UQ_{i}\phi_{2i} + \lambda_D D\tilde \phi_{2i} Q_i +\lambda_E E\tilde\phi_{2i}L_i
\end{equation}
where the $SU(2)$ indices are displayed. The usual hypercharge assignment leads to an anomaly-free model at scales below the $SU(3)$ breaking scale. We have put $E\equiv E_1$, $D\equiv D_{1}$ and
  $\tilde\phi_{2i}=\epsilon_{ji}\bar\phi_{2j}$. In particular we get
\begin{eqnarray}
\phi_2&=&(b,0)^T\nonumber \\
\tilde\phi_2&=&(0,b)^T\nonumber\\
\end{eqnarray}
along the flat direction.
This superpotential leads to masses for the fermions in a way similar to the supersymmetric standard model. The fermion mass scale is given by the vev of $b$, i. e. exponentially small compared to the $SU(3)$ breaking scale. 

Let us assume that the breaking of $SU(3)$ is set at the grand unified scale $10^{16}GeV$. The mass $M={{\pi R}\over 2}$ is also of order of the GUT scale,  the $SU(2)$ breaking can occur at a scale as low as the weak scale as long as the length of the fifth dimension is of order a few GUT lengths. More precisely relating the four and five dimensional gauge coupling constants $\alpha_5=2\pi R\alpha_4$ and using the  GUT value of the four dimensional coupling one finds
\begin{equation}
\alpha_5(4\pi T)^{-{1\over 2} }\sim 1
\end{equation}
In five dimensions  this implies that the model is not weakly coupled. This  suggests that the hierarchy problem is linked to the strong coupling regime of string theory. 
 From the point of view of low energy physics  the breaking of $SU(2)$ occurs at the weak scale due to a negative mass squared in the effective potential  for the Higgs field $b$. This effective potential  is of the sort that one has to postulate in the standard model. Moreover the superpotential $W$ leads to masses for the fermions as in the supersymmetric standard model. This may be a hint that extra dimensions with discrete fixed points  play a significant role in the breaking of the electroweak symmetry. 

We have presented  a simplified situation where a gauge symmetry is partially broken down to $SU(2)$ on one end point of a compactified fifth dimension. The second stage of symmetry breaking is induced by the presence of a massive Kaluza-Klein state propagating between the end points. It occurs on the other end point of the fifth dimension. The effective potential describing the self interaction of the Higgs field is similar to the one postulated in the standard model. In particular it possesses a negative mass term. Of course one would like to find a more realistic model where the compactification procedure is explicitly derived from string theory and the fermion superpotential calculated.
 This is left for future work. 
\vfill\eject
\centerline{\bf References}
\vskip 1 cm
\leftline{ [1] Nilles H. P. Phys. Rep. {\bf 110} 5.}
\vskip .2 cm
\leftline{ [2] Barbieri R., Ferrara S. and Savoy C. Phys. Lett. {\bf B119} (1982) 343.}
\vskip .2 cm
\leftline{ [3] Chamseddine A. H., Arnowitt R. and Nath P. Phys. Rev. Lett {\bf 49} (1982) 970.}
\vskip .2 cm
\leftline{ [4] Deredinger J. P. and Savoy C. Nucl. Phys. {\bf B237} (1984) 307.}
\vskip .2 cm 
\leftline{ [5] Cremmer E., Ferrara S., Kounnas C. and Nanopoulos D. V. Phys. Lett. {\bf B133} (1983) 61. }
\vskip .2 cm
\leftline{ [6] Ellis J., Kounnas C. and Nanopoulos D. V. Nucl. Phys. {\bf B 247} (1984) 373.}
\vskip .2 cm
\leftline{ [7] Witten E. Phys. Lett {\bf B155} (1985) 151.}
\vskip .2 cm
\leftline{ [8] Bailin D., Love A. and Thomas S. Nucl. Phys. {\bf B 273} (1986) 537.}
\vskip .2 cm
\leftline{ [9] Ferrara S., Kounnas C. and Porratti M. Phys. Lett. {\bf B181} (1986) 373.}
\vskip .2 cm
\leftline{ [10] Turok N. Phys. Rev. Lett. {\bf 76} (1996) 1015}
\vskip .2 cm
\leftline{ [11] Horava P. and Witten E. Nucl. Phys. {\bf B460} (1996) 506.}
\vskip .2 cm
\leftline{ [12] Antoniadis I., Ferrara S. and Taylor T. R. Nucl. Phys. {\bf B460} (1996) 489.}
\vskip .2 cm
\leftline{ [13] Cadavid A. C., Ceresole A., D'Auria R. and Ferrara S. "11d Supergravity}
\vskip .2 cm 
\leftline{ Compactified on Calabi-Yau Treefolds" hep-th/9506144.}
\vskip .2 cm
\leftline{ [14] Ferrara S., Khuri R. and Minasian R. "M-theory on a Calabi-Yau Manifold"}
\vskip .1 cm
 hep-th/9602102.
\vskip .2 cm
\leftline{[15] E. Dudas and C. Grojean "Four-Dimensional M-theory and Susy Breaking" hep-th/9704177.}
\vskip .2 cm
\leftline{[16] I. Antoniadis and M. Quiros "Susy Breaking in M-theory and} \vskip .1 cm
Gaugino Condensation" hep-th/9705037.
\vskip .2 cm
\leftline{ [17] Dixon L. ,  Harvey J. ,  Vafa C. and  Witten E. Nucl. Phys {\bf B261} (1985) 651, {\bf 274} (1986) 285.} 
\vskip .2 cm
\leftline{ [18]  Hamidi S. and  Vafa C. Nucl. Phys. {\bf B279} (1987) 465.}
\vskip .2 cm
\leftline{ [19]  Dixon L. , Friedan D.,  Martinec E. and Shenker S. Nucl. Phys. {\bf B282} (1987) 13.}
\vskip .2 cm
\leftline{ [20] Brax Ph. and Turok N.  "Exponentially Small Couplings Between Twisted Fields of }
\vskip .1 cm
Orbifold String Theories" hep-th/9706028.
\vskip .2 cm

\end{document}